\documentclass[12pt]{article}
\usepackage{amssymb,amsmath}
\usepackage[noblocks]{authblk}
\usepackage[top=0.75in, bottom=0.75in, left=0.75in, right=0.75in, dvips]{geometry}
\usepackage{caption}
\begin{document}
\textwidth 10.0in 
\textheight 9.0in 
\topmargin -0.60in
\title{Generators of Local Supersymmetry Transformation from First Class Constraints}
\author[1,2]{D.G.C. McKeon}
\affil[1] {Department of Applied Mathematics, The
University of Western Ontario, London, ON N6A 5B7, Canada} 
\affil[2] {Department of Mathematics and
Computer Science, Algoma University, Sault St.Marie, ON P6A
2G4, Canada}
\date{}
\maketitle

\maketitle
\noindent
email: dgmckeo2@uwo.ca

\begin{abstract}
We show how the generator of local supersymmetry transformations can be found from Fermionic first class constraints. This is done by adapting the approaches of Henneaux, Teitelboim and Zanelli and of Castellani that has been used to find the generator of gauge transformations from Bosonic first class constraints. We illustrate how a supersymmetric gauge generator can be found by considering the spinning particle. The invariances that we find  are not those presented in the original discussion of the spinning particle.
\end{abstract}

\section{Introduction}

Local gauge symmetries have long been associated with the presence of first class constraints that arise when applying the Dirac constraint formalism [1,2].  A precise expression for the generator of a gauge transformation in terms of these first class constraints can be found by either examining the invariances of the total action in phase space (the ``HTZ'' approach [3]) or by considering the equations of motion in phase space (the ``C'' approach [4]).  Local gauge transformations involving Bosonic gauge functions can be derived using this approach not only in ordinary Yang-Mills theory, but also in the first [5,6] and second order [7,8] Einstein-Hilbert (EH) action for $D > 2$ as well as the first order EH action when $D = 2$ [9].  In the latter case, a novel gauge transformation which is distinct from the manifest diffeomorphism transformation that is present has been uncovered through use of the gauge generator derived from the first class constraints.

In this paper, we will show that in addition to gauge transformations involving Bosonic gauge functions, gauge transformations involving Fermionic gauge functions can be derived by considering the first class constraints present in a model.  To illustrate this, we will first consider the spinning particle (``supergravity in $0 + 1$ dimensions'') [10,11].  In this model, the first class constraints that are present result in two distinct gauge transformations, one involving a Bosonic gauge function, the other a Fermionic gauge function.  These gauge transformations differ in certain respects from the gauge transformations that are manifest in the model. We also consider the extended spinning particle introduced in ref. [12].

\section{The Spinning Particle}

The action for the spinning particle with mass $m$ and $N = 1$ supersymmetry is [10,11]
\begin{align}
S &= \frac{1}{2} \int d\tau \bigg[ \eta_{\mu\nu} \bigg( \frac{\dot{\phi}^\mu (\tau) \dot{\phi}^\nu(\tau)}{e(\tau)} - i \psi^\mu(\tau) \dot{\psi}^\nu(\tau) - \frac{1}{e(\tau)} \chi(\tau)\dot{\phi}^\mu(\tau)\psi^\nu(\tau)\bigg) \nonumber \\
&\hspace{1cm} + \left( m^2 e(\tau) + i\psi_5(\tau) \dot{\psi}_5(\tau) - im\psi_5(\tau) \chi(\tau)\right)\bigg] 
\end{align}
where $\eta_{\mu\nu}$ is the ``target space'' metric, the fields $\phi^\mu$ and $e$ are Bosonic and $\psi^\mu$, $\psi_5$ and $\chi$ are Fermionic (Grassmann). This actions possesses the manifest ``diffeomorphism'' invariance 
\[ \delta \phi^\mu = f\dot{\phi}^\mu, \quad
\delta e  = \dot{f} e + e\dot{f}, \quad
\delta \chi  = \dot{f}\chi + \chi \dot{f}, \quad
\delta \psi^\mu = f\dot{\psi}^\mu, \quad \delta\psi_5 = f\dot{\psi}_5 \eqno(2a-e)
\]
and the manifest $N = 1$ supersymmetry [11]
\[
\delta \phi^\mu = i\alpha\psi^\mu, \quad \delta\psi^\mu = \frac{\alpha}{e} \left(\dot{\phi}^\mu - \frac{i}{2} \chi \psi^\mu\right),\quad
\delta e = i\alpha \chi, \quad \delta \chi = 2\dot{\alpha} \nonumber \]
\[ \delta \psi_5 = m\alpha + \frac{i}{me} \alpha \psi_5 \left(\dot{\psi}_5 - \frac{1}{2} m\chi\right).\eqno(3a-e)\]
The parameters $f(\tau)$, $\alpha(\tau)$ are Bosonic and Fermionic respectively.

We will now perform a full canonical analysis of the action $S$ of eq. (1) and show how the first class constraints can be used to find the generator of gauge transformations that leave $S$ invariant.  These will be compared with the manifest invariances of eqs. (2,3).

The canonical momenta following from $S = \int d\tau L$ are 
\[ \hspace{-3cm}p_\mu = \frac{\partial L}{\partial \dot{\phi}^\mu} = \frac{1}{e} \left( \dot{\phi}_\mu - \frac{i}{2} \chi \psi_\mu\right)\,\,, \quad p_e = \frac{\partial L}{\partial \dot{e}} = 0 \nonumber \]
\[ \pi_\mu = \frac{\partial L}{\partial \dot{\psi}^\mu} = \frac{i}{2} \psi_\mu\,\,, \quad 
\pi_5 = \frac{\partial L}{\partial \dot{\psi}_5} = -\frac{i}{2} \pi_5\,\,, \quad \pi_\chi = \frac{\partial L}{\partial \dot{\chi}} = 0,  \eqno(4a-e)\] 
so that the canonical Hamiltonian is 
\[ H_c = \frac{e}{2} (p^2 - m^2) + \frac{i}{2} \chi (p \cdot \psi - m\psi_5 ).\eqno(5) \]
The primary constraints of eqs. (4b,e) yield secondary constraints as 
\[\hspace{-1.2cm} \left\{ p_e, H_c \right\} = - \frac{1}{2} (p^2 - m^2) \eqno(6a) \]
\[ \left\{\pi_\chi, H_c \right\} = - \frac{i}{2} (p \cdot \psi -  m \psi_5). \eqno(6b) \]
In addition, the constraints of eqs. (4c,d) are second class as 
\[ \left\{\pi_\mu - \frac{i}{2} \pi_\mu, \pi_\nu - \frac{i}{2} \pi_\nu\right\} = i\eta_{\mu\nu}\,,\quad 
\left\{\pi_5 + \frac{i}{2} \psi_5, \pi_5 + \frac{i}{2} \psi_5\right\} = -i, \eqno(7a,b) \]
so that the Dirac Brackets are given by 
\[\left\{A,B\right\}^* = \left\{A,B\right\} + i \left[ \left\{ A, \psi_\mu - \frac{i}{2} \psi_\mu \right\} \left\{ \psi^\mu - \frac{i}{2} \psi^\mu, B\right\}\right. \nonumber \]
\[ \left. -\left\{ A, \pi_5 + \frac{i}{2} \psi_5 \right\} \left\{ \pi_5 + \frac{i}{2} \psi_5, B\right\}\right] \eqno(8) \]
for dynamical variables $A,B$.  In particular, it follows that 
\[ \left\{ \psi_\mu, \psi_\nu \right\}^* = i\eta_{\mu\nu}\,\, ,\quad 
\left\{ \psi_5, \psi_5 \right\}^* = -i. \eqno(9a,b) \]
The constraints of eqs. (4b,e; 6a,b) are all first class as 
\[ \left\{ p \cdot \psi - m  \psi_5, p \cdot \psi - m\psi_5 \right\}^* = i(p^2 - m^2).  \eqno(10) \]
We can show that the HTZ formalism of ref. [3], originally introduced to find the generator of gauge transformations for systems with only Bosonic degrees of freedom, can be employed to deal with systems with both Bosonic and Fermionic degrees of freedom.  In particular, for the model of eq. (1), we have a gauge generator $G$ given by 
\[ G = B_1 p_e + B_2(p^2 - m^2) + i F_1 \pi_\chi + i F_2 (p\cdot \psi - m\psi_5 ) \eqno(11) \]
where $B_i$ and $F_i$ are Bosonic and Fermionic gauge functions respectively.  With a total Hamiltonian
\[ H_T = H_c + \lambda_e p_e + i\lambda_\chi \pi_\chi \eqno(12) \]
($\lambda_e$ and $\lambda_\chi$ are Lagrange multipliers), then by ref. [3], we have the equation
\[ \dot{B}_1 p_e + \dot{B}_2 (p^2 - m^2) + i\dot{F}_1 \pi_\chi + i \dot{F}_2 (p \cdot \psi - m\psi_5) \nonumber \]
\[ + \left\{ G, H_T\right\}^* - \delta \lambda_e p_e - i\delta \lambda_e \pi_\chi = 0. \eqno(13) \]
By considering the coefficients of the constraints $p^2 - m^2 = 0$ and $p \cdot \psi - m\psi_5 = 0$ respectively, it follows from eq. (13) that 
\[ B_1 = 2\dot{B}_2 + i F_2 \chi \eqno(14a) \]
\[ F_1 = -2i\dot{F}_2 \eqno(14b) \]
so that if $B_2 = B$, $F_2 = F$
\[ G = (2 \dot{B} + i F\chi) p_e + B(p^2 - m^2) + 2 \dot{F} \pi_\chi + iF(p \cdot \psi - m\psi_5). \eqno(15) \]
We note that the $C$ formalism of ref. [4] can also be used to find $G$.  With there being two generators of constraints, the form of $G$ is 
\[ G = \epsilon(\tau) G_0 + \dot{\epsilon}(\tau) G_1 \eqno(16) \]
where [4] 
\[ G_1 = \mathrm{primary \;first\; class\; constraint} (P1) \eqno(17a) \]
\[ G_0 + \left\{G_1, H_T\right\}^* = P1 \eqno(17b) \]
\[ \left\{ G_0, H_T \right\}^* = P1  .\eqno(17c) \]
Satisfying eq. (17a) by taking
\[ G_1 = p_e\eqno(18a) \]
we see that eq. (17b) leads to 
\[ G_0 = \frac{1}{2} (p^2 - m^2) + \alpha_e p_e + \alpha_\chi p_\chi ;\eqno(18b) \]
by eq. (17c), the Lagrange multipliers $\alpha_e$ and $\alpha_\chi$ must satisfy
\[ \alpha_e = \alpha_\chi = 0 \eqno(18c) \]
and so by eq. (16) 
\[ G_A = \frac{\epsilon_A}{2} (p^2 - m^2) + \dot{\epsilon}_A p_e .\eqno(19) \]
Similarly, if in eq. (17a) we take
\[ G_1 = \pi_\chi \eqno(20a) \]
then by eq. (17b)
\[ G_0 = \frac{i}{2} \left( p \cdot \psi - m\psi_5 \right) + \beta_e p_e + \beta_\chi \pi_\chi \eqno(20b) \]
and with eq. (17c) we find that
\[ \beta_\chi = 0\,, \quad \beta_e = \frac{i\chi}{2} \eqno(20c) \]
so that from eqs. (16,20) 
\[ G_B = \epsilon_B \left( \frac{i}{2}\left( p \cdot \psi - m \psi_5 \right) + \frac{i\chi}{2} p_e\right) + \dot{\epsilon}_B\, \pi_\chi .\eqno(21) \]
The full generator $G_A + G_B$ is identical to $G$ of eq. (15) if $2B = \epsilon_A$, $2F = \epsilon_B$.

With the gauge transformation in a dynamical variable $A$ being given by $\delta A = \left\{ A, G \right\}^*$, we find that 
\[\hspace{-3.5cm} \delta \phi^\mu = 2Bp^\mu + iF\psi^\mu\nonumber \]
or, by eq. (4a) 
\[ = \frac{2B}{e} \left(\dot{\phi}^\mu - \frac{i}{2} \chi\psi^\mu\right) + iF\psi^\mu ,\eqno(22a) \]
\[ \delta e = 2 \dot{B} + iF\chi \eqno(22b) \]
\[\hspace{-1cm} \delta\chi = 2\dot{F}\eqno(22c) \]
\[\hspace{-1cm} \delta\psi^\mu = Fp^\mu \eqno(22d) \]
\[\hspace{2.1cm} = \frac{F}{e} \left( \dot{\phi}^\mu - \frac{i}{2} \chi\psi^\mu \right) \eqno(22d) \]
\[ \hspace{-1cm}\delta\psi_5 = mF. \eqno(22e) \]
Eq. (22) is not identical to eqs. (2) and (3).  First of all, the Fermionic fields $\chi$, $\psi^\mu$ and $\psi_5$ do not change if $F = 0$ where as they do change under the transformation of eq. (2); in the limit $\chi = \psi^\mu = \psi_5 = 0$, eqs. (2) and (22) are identical only if $2B = ef$ so that $B$ acquires a dependence on the dynamical field $e$.

Secondly, the Fermionic portion of eq. (22) is identical to eq. (3) when $F = \alpha$ only if we ignore the contribution of the second term in $\delta\psi_5$ appearing in eq. (3e).  However, one can 
establish that this term by itself leaves the action invariant and therefore need not participate in the Fermionic portion of the transformation of eq. (22).

It is also worth noting that if gauge parameters $B_i$ and $F_i$ are associated with generator $G_i$ in eq. (15), then
\[ \left\{ G_i, G_j \right\}^* = 2 \frac{d}{d\tau} (i F_iF_j)p_e + (iF_1F_2)(p^2-m^2).\eqno(23)\]
Consequently, the Dirac Bracket of two generators $G_i$ and $G_j$ is itself a purely Bosonic generator with gauge parameter $B = iF_iF_j$. The gauge transformations given in refs. [10,11] do not obey the algebra implied by eq. (23).

We now will consider the generator of supersymmetry transformations for the spinning string with $N > 1$ supersymmetry.  The action for this model is [12]
\[ S = \frac{1}{2}  \int d\tau \left[ \frac{1}{e} \left(\dot{\phi}^\mu - \frac{i}{2} \chi_i \psi_i^\mu\right) \left(\dot{\phi}^\nu  - \frac{i}{2} \chi_j\psi_j^\nu \right) \right. \eqno(24) \]
\[\left. \hspace{2cm} - \frac{i}{2} \psi_i^\mu \dot{\psi}_i^\nu - \frac{i}{2} f_{ij}\psi_i^\mu \psi_j^\nu \right] \eta_{\mu\nu}\nonumber \]
where there are now $N$ Fermionic fields $\psi_i^\mu$, $\chi_i\,(i = 1 \ldots N)$ and $f_{ij}$ is an antisymmetric Boson field.  The momenta associated with $(\phi^\mu,\, e,\, \psi_i^\mu,\,\chi_i ,\, f_{ij})$ are now respectively
\[ p^\mu = \frac{1}{e}\left( \dot{\phi}^\mu - \frac{i}{2} \chi_i \psi_i^\mu\right)\,, \quad p_e = 0\,, \quad \pi_{i\mu} = \frac{i}{2} \psi_{i\mu}\,, \quad \pi_{i\chi} = 0\,, \quad p_{ij} = 0 \eqno(25a-e) \]
respectively and thus the canonical Hamiltonian is given by 
\[ H_c = \frac{e}{2} p^2 + \frac{i}{2} \chi_i p \cdot \psi_i + \frac{i}{2} f_{ij} \psi_i \cdot \psi_j .\eqno(26) \]
The second class constraints of eq. (25c) lead to the Dirac Bracket
\[\left\{ A, B \right\}^* = \left\{ A, B \right\} + i \delta_{ij} \eta_{\mu\nu} \left\{ A, \pi_i^\mu - \frac{i}{2} \psi_i^\mu \right\}
\left\{  \pi_j^\nu - \frac{i}{2} \psi_j^\nu , B \right\}.\eqno(27) \]
The primary constraints of eqs. (25b,d,e) lead respectively to the secondary constraints
\[ \left\{ p_e, H_c \right\}^* = - \frac{1}{2} p^2\,, \quad 
\left\{ \pi_{i\chi}, H_c \right\}^* = - \frac{i}{2} p \cdot \psi_i\,, \quad
\left\{ p_{ij}, H_c \right\}^* = - \frac{i}{2} \psi_i \cdot \psi_j .
\eqno(28a-c) \]
Since
\[\hspace{-5.02cm} \left\{ p \cdot \psi_i, H_c \right\}^* = \frac{1}{2} \chi_i p^2 - f_{ij} p \cdot \psi_j \eqno(29a) \]
\[ \left\{\psi_i \cdot \psi_j, H_c \right\}^* = \frac{1}{2} \left( \chi_i p \cdot \psi_j -  \chi_j p \cdot \psi_i\right) + \left( f_{im} \psi_j - f_{jm} \psi_i\right) \cdot \psi_m\eqno(29b) \]
and
\[ \left\{\psi_i \cdot \psi_j, \psi_k \cdot \psi_\ell \right\}^* = -i \left( \delta_{ik} \psi_j \cdot \psi_\ell - \delta_{i\ell} \psi_j \cdot \psi_k \right. \nonumber \]
\[ \hspace{2cm}\left. + \delta_{j\ell} \psi_i \cdot \psi_k - \delta_{jk} \psi_i \cdot \psi_\ell \right)\eqno(30a)\]
\[\hspace{-3cm} \left\{ p \cdot \psi_i, p \cdot \psi_j\right\}^* = ip^2 \delta_{ij} \eqno(30b) \]
\[\hspace{-.35cm} \left\{ p \cdot \psi_i, \psi_k  \cdot \psi_\ell \right\}^* = i \left( \delta_{ik} p \cdot \psi_\ell - \delta_{i\ell} p \cdot \psi_k\right) \eqno(30c) \]
we see that there are no further constraints and that all constraints other than eq. (25c) are first class.

The gauge generator resulting from these first class constraints is now of the form
\[ G = B_1 p_e + B_2p^2 + \overline{B}^{\,ij}_1 p_{ij}
 + i\overline{B}^{\,ij}_2 \psi_i \cdot \psi_j
 + iF_1^i \pi_{i\chi} + iF_2^i p\cdot \psi_i  \,.  \eqno(31) \]
The HTZ formalism can now be applied; in analogy with eq. (14) we find that
\[\hspace{-1.8cm} B_1 = 2 \left( \dot{B}_2 + \frac{i}{2} F_2^i \chi_i\right) \eqno(32a) \]
\[ \overline{B}^{\,ij}_1 = 2 \left( \dot{\overline{B}}^{\,ij}_2
+ \overline{B}^{\,ik}_2 f_{jk} - \overline{B}^{\,jk}_2 f_{ik}\right) \eqno(32b) \]
\[\hspace{-.1cm} F_1^i = -2i\left( \dot{F}_2^i - \overline{B}^{\,ij}_2 \chi_j + f_{ij} F_2^j \right) \eqno(32c) \]
so that with $F_2 \equiv F$, $B_2 = B$, $\overline{B}_2 = \overline{B}_2$
\[ G = 2\left( \dot{B} + \frac{i}{2} F^i \chi_i \right) p_e + Bp^2 + 2 \left( \dot{\overline{B}}^{\,ij} + \overline{B}^{ik} f_{jk}  - \overline{B}^{jk} f_{ik} \right) p_{ij}\nonumber \]
\[ + i\overline{B}^{\,ij} \psi_i \cdot \psi_k + 2 \left( \dot{F}^i - \overline{B}^{\,ij} \chi_j + f_{ij} F^j \right)\pi_{i\chi}  + iF^i p \cdot 
\psi_i .\eqno(33) \]
The gauge transformation of any dynamical variable $A$ is now given by $\delta A = \left\{ A,G \right\}^*$; it is evidently not identical to the transformations given in ref. [12] or the $N = 2$ limit discussed in ref. [11].
\section{Discussion}

We have shown how the Dirac constraint formalism can be adapted to find in generator of local gauge invariances that are supersymmetry transformations.  We have demonstrated this by finding the gauge generator arising from first class constraints for the spinning particle with $N \geq 1$ supersymmetry whose actions were originally given in refs. [10,11,12].

It would be of interest to apply this approach to other systems in which a local supersymmetry is manifest in order to see if the first class constraints present can be employed to find these supersymmetries.  Among such systems are the spinning string [13], the spinning membrane [14], the super particle [15] and superstring [16].  In these latter two systems, it was first noticed that there is a global supersymmetry, but subsequently a local supersymmetry was uncovered [17].  We would like to see if this so-called ``$\kappa$ symmetry'' is a consequence of the presence of first class constraints.  Finally, it would be worth examining the relationship between the presence of first class constraints and supergravity in such models as these of ref. [18], especially since the local supersymmetry transformations are only invariances of the action on the mass shell if there are no auxiliary fields.

It is possible to make use of the gauge generator to choose a covariant gauge when using the path integral derived from the canonical quantization procedure [19].  We hope to apply this approach to the quantization of supergravity theories.

We would also like to see if there are any models with Fermionic first class constraints which imply gauge transformations that are not obviously related to a local supersymmetry.
 
\section*{Acknowledgements}
We would like to thank S. Kuzmin and N. Kiriushcheva for useful  conversations. R. Macleod made a helpful comment.

\section*{Appendix}

We use a ``left derivative'' for Grassmann variables; if $\theta_I$ are Grassmann, then
\[  \frac{\partial}{\partial\theta_A} (\theta_B\theta_C) = \delta_{AB} \theta_C - \delta_{AC}\theta_B,\quad \frac{d}{d t}F(\theta(t)) = \dot{\theta}(t) F^\prime (\theta(t)). \eqno(A.1a,b) \]
If $F_i(B_i)$ are Grassmann odd (even) quantities and in phase space $(q_i, p_i)$ $((\psi_i, \pi_i))$ are ordinary (Grassmann) conjugate pairs, then we define the Poisson Brackets by:
\[ \left\{ B_1, B_2 \right\} = \left( B_{1,q} B_{2,p} - B_{1,p}B_{2,q} \right) + \left( B_{1,\psi} B_{2,\pi} - B_{2,\psi}B_{1,\pi} \right) \eqno(A.2a)\]
\[\hspace{-1cm} \left\{ F,B \right\} = \left( F_{,q} B_{,p} - B_{,q}F_{,p} \right) - \left( F_{,\psi} B_{,\pi} + B_{,\psi}F_{,\pi} \right) \eqno(A.2b)\]
\[\hspace{-1cm} \left\{ B,F \right\} = \left( B_{,q} F_{,p} - F_{,q}B_{,p} \right) + \left( F_{,\psi} B_{,\pi} + B_{,\psi}F_{,\pi} \right) \eqno(A.2c)\]
\[\hspace{-.2cm} \left\{ F_1,F_2 \right\} = \left( F_{1,q} F_{2,p} + F_{2,q}F_{1,p} \right) - \left( F_{1,\psi} F_{2,\pi} + F_{2,\psi}F_{1,\pi} \right) \eqno(A.2d)\]
where $B_{1,q} B_{2,p} = \displaystyle{\sum_{i}} \frac{\partial B_1}{\partial q_i} \frac{\partial B_2}{\partial p_i}$ etc.

If $L = L(q_i,\psi_i)$ is the Lagrangian, then the Hamiltonian is
\[ H = \dot{q}_i p_i + \dot{\psi}_i \pi_i - L .\eqno(A.3) \]


\begin{thebibliography}{99}
\bibitem{1} P.A.M. Dirac, ``Lectures on Quantum Mechanics'' (Dover, Mineoloa, 2001).
\bibitem{2} P.A.M. Dirac, \textit{Can. J. Math.} \textbf{2}, 129 (1950); \textbf{3} 1 (1951).
\bibitem{3} M. Henneaux, C. Teitelboim and J. Zanelli, \textit{Nucl. Phys.} \textbf{B332}, 169 (1990).
\bibitem{4} L. Castellani, \textit{Ann. Phys.} \textbf{142}, 357 (1982).
\bibitem{5} D.G.C. McKeon, \textit{Int. J. Mod. Phys} \textbf{A25}, 3453 (2010).
\bibitem{6} N. Kiriushcheva and S.V. Kuzmin, \textit{Eur. J. Phys.} \textbf{C70}, 389 (2010). 
\bibitem{7} N. Kiriushcheva, S.V. Kuzmin, C. Racknor and S.R. Valluri, \textit{Phys. Lett.} \textbf{A372}, 5101 (2008).
\bibitem{8}  N. Kiriushcheva and S.V. Kuzmin, \textit{Central Eur. J. Phys.} \textbf{9}, 576 (2011). 
\bibitem{9}  N. Kiriushcheva and S.V. Kuzmin, \textit{Mod. Phys. Lett.} \textbf{A20}, 1895 (2005). 
\bibitem{10} L. Brink, S. Deser, B. Zumino, P.DiVecchia and P. Howe,  \textit{Phys. Lett.} \textbf{64B}, 435 (1976).
\bibitem{11} L. Brink, P. DiVecchia and P. Howe,  \textit{Nucl. Phys.} \textbf{B118}, 76 (1977).
\bibitem{12} P. Howe, S. Penati, M. Pernici and P. Townsend, \textit{Phys. Lett.} \textbf{215B}, 555 (1988).
\bibitem{13} S. Deser and B. Zumino, \textit{Phys. Lett.} \textbf{65B}, 369 (1976).\\
L. Brink, P. DiVecchia and P. Howe, \textit{Phys. Lett.} \textbf{65B}, 471 (1976).
\bibitem{14} P. Howe and R. Tucker,  \textit{J. Phys.} \textbf{A10}, L155 (1977).
\bibitem{15} L. Brink and J. Schwarz, \textit{Phys. Lett.} \textbf{100B}, 310 (1981).
\bibitem{16} M. Green and J. Schwarz, \textit{Phys. Lett.} \textbf{109B}, 449 (1982); \textbf{136B}, 367 (1984).
\bibitem{17} W. Siegel, \textit{Phys. Lett.} \textbf{128B}, 397 (1983).
\bibitem{18} D.Z. Freedman, P. van Nieuwenhuizen and S. Ferrara, \textit{Phys. Rev.} \textbf{D13}, 3214 (1976).\\
S. Deser and B. Zumino, \textit{Phys. Lett.} \textbf{62B}, 335 (1976).
\bibitem{19} D.G.C. McKeon, \textit{Can. J. Phys.} (in press).
\end{thebibliography}
\end{document}